\documentclass[runningheads]{llncs}
\usepackage[T1]{fontenc}
\usepackage{lscape}
\usepackage{graphicx}
\usepackage{amsmath} 
\usepackage{algpseudocode}
\usepackage{algorithm}
\usepackage{color}
\begin{document}
\title{On-demand Cold Start Frequency Reduction with Off-Policy Reinforcement Learning in Serverless Computing}
\titlerunning{On-demand Cold Start Frequency Reduction in Serverless Computing}
\author{Siddharth Agarwal
\and
Maria A. Rodriguez
\and
Rajkumar Buyya}
\authorrunning{S. Agarwal, M.A. Rodriguez and R. Buyya}
\institute{Cloud Computing and Distributed Systems Laboratory\\
School of Computing and Information Systems\\
The University of Melbourne, Australia\\
\email{siddhartha@student.unimelb.edu.au}\\
\email{\{maria.read,rbuyya\}@unimelb.edu.au} \\
}
\maketitle              % typeset the header of the contribution
\begin{abstract}
Function-as-a-Service (FaaS) is a cloud computing paradigm offering an event-driven execution model to applications. It features ‘serverless’ attributes by eliminating resource management responsibilities from developers, and offers transparent and on-demand scalability of applications. To provide seamless on-demand scalability, new function instances are prepared to serve the incoming workload in the absence or unavailability of function instances. However, FaaS platforms are known to suffer from \textit{cold starts}, where this function provisioning process introduces a non-negligible delay in function response and reduces the end-user experience. Therefore, the presented work focuses on reducing the frequent, on-demand cold starts on the platform by using Reinforcement Learning(RL). The proposed approach uses model-free Q-learning that consider function metrics such as CPU utilization, existing function instances, and response failure rate, to proactively initialize functions, in advance, based on the expected demand. The proposed solution is implemented on Kubeless and evaluated using an open-source function invocation trace applied to a \textit{matrix multiplication} function. The evaluation results demonstrate a favourable performance of the RL-based agent when compared to Kubeless' default policy and a function keep-alive policy by improving throughput by up to 8.81\% and reducing computation load and resource wastage by up to 55\% and 37\%, respectively, that is a direct outcome of reduced cold starts.

\end{abstract}
\section{Introduction}
\label{introduction}

In cloud computing, a serverless deployment model removes the burden of managing and provisioning resources from the developers, allowing them to focus solely on the application development process. The term \textit{serverless}, interchangeably used with Function-as-a-Service (FaaS), does not imply an absence of servers, but instead, accentuates delegating the responsibility of complex resource management tasks to cloud service providers (CSP) \cite{lloyd2018serverless}, \cite{jonas2019cloud}. The FaaS paradigm puts forward an event-driven, serverless computing model with fine-grained pay-per-use pricing where resources are billed based on their actual service time. Functions (i.e., a fragment of code containing business logic) are designed to scale on demand; they are stateless, short-lived, and run on lightweight containers or virtual machines (VMs) in response to a triggering event. Such an abstraction increases agility in application development, offering lower administrative and ownership costs. 
The FaaS model has attracted a wide range of applications such as IoT services, REST APIs, stream processing and prediction services, which have strict availability and quality of service requirements in terms of response time. Conceptually, the FaaS model is designed to spin a new function instance for each demand request and shut down the instance after service \cite{jonas2019cloud}. However, in practice, commercial FaaS offerings like AWS Lambda, Azure Functions, and Google Cloud Function may choose to re-use a function instance or keep the instance running for a limited time to serve subsequent requests \cite{vahidinia2020cold}. Some open source serverless frameworks like Kubeless \cite{kubeless} and Knative, have similar implementations to re-use an instance of a function to serve subsequent requests.

An increase in workload demand leads to an instantiation process involving the creation of new function containers and the initialisation of the function's environment within those containers, after which incoming requests are served. Such a process usually requires downloading the client code, setting up code dependencies and the runtime environment, setting up container networking, and finally executing the code to handle the incoming request. Hence, instantiating a function's container introduces a non-negligible time latency, known as \emph{cold start}, and gives rise to a challenge for serverless platforms \cite{xu2019adaptive}, \cite{lin2019mitigating}, \cite{bermbach2020using}, \cite{shafiei2019serverless}. Some application-specific factors such as programming language, runtime environment and code deployment size as well as function requirements like CPU and memory, affect the cold start of a function \cite{shafiei2019serverless}, \cite{mohan2019agile}, \cite{manner2018cold}, \cite{solaiman2020wlec}.
To automate the process of creating new function instances and reusing existing ones, serverless frameworks usually rely on resource-based (CPU or memory) horizontal scaling, known as horizontal pod auto-scaling (HPA) in Kubernetes-based frameworks like Kubeless, to respond to incoming requests. Resource-based scaling policies implement a reactive approach and instantiate new functions only when resource usage rises above a pre-defined threshold, thus leading to cold start latencies and an increase in the number of unsuccessful requests.

Threshold-based scaling decisions fail to consider factors like varying application load and platform throughput and hence, pose an opportunity to explore dynamic techniques that analyse these factors to address cold starts. This work presents a model-free Q-learning agent to exploit resource utilization, available function instances, and platform response failure rate to reduce the number of cold starts for CPU-intensive serverless functions. We define a reward function for the RL agent to dynamically establish the required number of function instances for a given workload demand based on expected average CPU utilisation and response failure rate. The RL-based agent interacts with the serverless environment by performing scaling actions and learns through trial and error during multiple iterations. The agent receives delayed feedback, either positive or negative, based upon the observed state, and consequently learns the appropriate number of function instances to fit the workload demand. This strategy uses no prior knowledge about the environment, demand pattern or workload, and dynamically adjusts to the changes for preparing required functions in advance to reduce cold starts. 
The proposed work scales the number of function instances by proactively estimating the number of functions that are needed to serve incoming workload to reduce the frequent cold starts. It utilizes a practical workload of matrix multiplication involved in an image processing task, serving as a sample real-world function request pattern \cite{shahrad2020serverless}, and formally presents the cold start as an optimisation problem. Also, we structure the Q-learning components around the function metrics such as average CPU utilisation and response failure rate and evaluate our approach against the default resource-based policy and commercially accepted function keep-alive technique. 

In summary, the key contributions of our work are:
\begin{enumerate}
\item We analyze function resource metrics such as CPU utilization, available instances, and the proportion of unsuccessful responses to propose a Q-learning model that dynamically analyses the application request pattern and improves function throughput by reducing frequent cold starts on the platform. 
\item We present a brief overview of explored solutions to address function cold starts and highlight the differences between contrasting approaches to the proposed agent.
\item We perform our experiments on a real-world system setup and evaluate the proposed RL-based agent against the default resource-based policy and a baseline keep-alive technique.
\end{enumerate}

The rest of the paper is organised as follows. Section \ref{section2} highlights related research studies. In Section \ref{section3}, we present the system model and formulate the problem statement. Section \ref{section4} outlines the proposed agent’s workflow and describes the implementation hypothesis and assumptions. In Section \ref{section5}, we evaluate our technique with the baseline approach and highlight training results and discuss about performance in Section \ref{section6}. Section \ref{section7} highlights future research direction and Section \ref{section8} summarises and concludes the paper.

\section{Related Work}
\label{section2}

In this section, we briefly discuss about the Function-as-a-Service paradigm in serverless computing and elaborate on the current function cold start mitigation techniques and approaches.

\subsection{Serverless Computing or Function-as-a-Service}

Serverless computing offers a cloud service model wherein the server management or resource management responsibility lies with the CSP. In \cite{jonas2019cloud}, the authors discussed the potential of this new, less complex computing model introduced by Amazon in 2014. They explain a function-based, serverless commercial offering of AWS Lambda, i.e., the Function-as-a-Service platform. They highlighted three primary differences between traditional cloud computing and serverless computing as follows: decoupled computation and storage, code execution without resource management, and paying in proportion to the resources used.The research posits that the serverless or FaaS model promotes business growth, making the use of the cloud easier.

Baldini \emph{et al.} \cite{baldini2017serverless} introduced the emerging paradigm of FaaS as an application development architecture that allows the execution of a piece of code in the cloud without control over underlying resources. They identified containers and the emergence of microservices architecture as the promoter of FaaS model in serverless. They used FaaS and serverless interchangeably and defines it as a ‘stripped down’ programming model that executes stateless functions as its deployment unit.

Since the inception of serverless computing, there have been many commercial and open-source offerings such as AWS Lambda, Microsoft Azure Functions, Google Cloud Functions, Fission, and OpenWhisk. These platforms represent FaaS as an emerging technology but Hellerstein \emph{et al.} \cite{hellerstein2018serverless} put together gaps that furnish serverless as a bad fit for cloud innovations. The authors criticized the current developments of cloud computing and state that the potential of cloud resources is yet to be harnessed. On the contrary, the work in \cite{shafiei2019serverless} argued that serverless offerings are economical and affordable as they remove the responsibility of resource management and complexity of deployments from consumers. It presented the opportunities offered by multiple FaaS offerings and gives an overview of other existing challenges and indicates potential approaches for future work. 

A Microsoft work \cite{rosenbaum} estimated that there will be near 500 million new applications in the subsequent 5 years, and it would be difficult for the current development models to support such large expansions. Another recent study by Datadog \cite{datadog}, publishes that over 70\% organisations using AWS cloud services, 50\% organisations using Microsoft Azure services and Google Cloud platform have adopted serverless computing into their architectures. FaaS is designed to increase development agility, reduce the cost of ownership, and decrease overheads related to servers and other cloud resources. The term 'serverless' has been in the industry since the introduction of Backend-as-a-Service (BaaS). Despite the serverless benefits, FaaS experiences two major challenges, which are categorized as (i) system-level and (ii) programming and DevOps challenges \cite{jonas2019cloud}, \cite{baldini2017serverless}, \cite{rosenbaum}. The former identifies the cost of services, security, resource limits, and cold start while scaling, and the latter focuses on tools and IDEs, deployment, statelessness, and code granularity in the serverless model.

\subsection{Function Cold Start and Mitigation}

Researchers in \cite{vahidinia2020cold} described function cold start as the time taken to execute a function. This process involves assigning a container to a function, accessing the code package and copying the function image, loading the image into memory, unpacking it, and executing the function handler. It broadly classified the approaches to deal with function cold start in, environment optimization, and pinging.
The former approach acts either by reducing container preparation time or decreasing the delay in loading function libraries, while the latter technique continuously monitors the functions and periodically pings them to keep the instances warm or running.

An adaptive container warm-up technique to reduce the cold start latency and a container pool strategy to reduce resource wastage is introduced in \cite{xu2019adaptive}. The proposed solution leverages a Long-Short Term Memory (LSTM) network to predict function invocation times and non-first functions in a chain to keep a warm queue of function containers ready. Although both the discussed techniques work in synchronization, the first function in the chain suffered from a cold start. 

The research in \cite{lin2019mitigating} explained platform-dependent overheads like pod provisioning and application implementation-dependent overheads. It presented a pool-based pre-warmed container technique, marked with selector ‘app-label’ to deal with the function cold start problem. To tackle the incoming demand, a container pool is checked first for existing pre-warmed containers, or the platform requests new containers as per the demand.

Another study \cite{mahajan2019exploiting} exploited the data similarity to reduce the function cold start. It criticized the current container deployment technique of pulling new container images from the storage bucket and introduced a live container migration over a peer-to-peer network. Similarly, \cite{bermbach2020using} aimed to reduce the number of cold start occurrences by utilizing the function composition knowledge. It presented an application-side solution based on lightweight middleware. This middleware enable the developers to control the frequency of cold start by treating the FaaS platform as a black box.

Based on the investigation in \cite{mohan2019agile}, network creation and initialization were found to be the prime contributors to the cold start latency. The study expressed that cold starts are caused due to work and wait times involved in various set-up processes like initializing networking elements. The study explained the stages of the container lifecycle and states that the clean-up stage demands cycles from the underlying containerization daemon, hindering other processes. Therefore, a paused container pool manager is proposed to pre-create a network for function containers and attach the new function containers to configured IP and network when required.

Some studies \cite{shafiei2019serverless}, \cite{manner2018cold}, \cite{hellerstein2018serverless} have identified significant factors that affect the cold start of a function. These include runtime environment, CPU and memory requirements, code dependency setting, workload concurrency, and container networking requirements. Most works \cite{shilkov2021comparison}, \cite{lee2018evaluation}, \cite{wang2018peeking}, \cite{mcgrath2017serverless}, \cite{lynn2017preliminary} focus on commercial FaaS platforms like AWS Lambda, Azure Functions, Google Cloud Functions and fall short to evaluate open source serverless platforms like OpenLambda, Fission, Kubeless, etc. Very few studies \cite{solaiman2020wlec}, \cite{santos2019towards}, \cite{mohanty2018evaluation} have successfully performed analysis on an open-source serverless platform and provided possible solution by targeting the container level fine-grained control of the platform.

Recent research works \cite{schuler2021ai}, \cite{vahidinia2022mitigating}, \cite{benedetti2022reinforcement}, \cite{agarwal2021reinforcement} introduce the paradigm of RL to the FaaS platforms in different ways. \cite{schuler2021ai} focuses on request-based provisioning of VMs or containers on the Knative platform. The authors demonstrated a correlation between latency and throughput with function concurrency levels and thus propose a Q-Learning model to determine the optimal concurrency level of a function for a single workload. \cite{vahidinia2022mitigating} proposed a two-layer adaptive approach,an RL algorithm to predict the best idle-container window, and an LSTM network to predict future invocation times to keep the pre-warmed containers ready.  

The study demonstrated the advantages of the proposed solution on the OpenWhisk platform using a simple HTTP-based workload and a synthetic demand pattern. Another research \cite{benedetti2022reinforcement} focused on resource-based scaling configuration (CPU utilisation) of OpenFaaS and adjusts the HPA settings using an RL-based agent. They assumed a serverless-edge application scenario and a synthetic demand pattern for the experimentation and present their preliminary findings based on latency as SLA. 

Agarwal \emph{et al.} \cite{agarwal2021reinforcement} introduced the idea of Q-learning to ascertain the appropriate amount of resources to reduce frequent cold starts. The authors shared the preliminary training results with an attempt to show the applicability of reinforcement learning to the serverless environment. They utilised the platform exposed resource metrics to experiment with a synthetic workload trace, i.e. Fibonacci series calculation, to simulate a compute-intensive application and predict the required resources.

Our proposed work introduces a Q-Learning strategy to reduce frequent cold starts in the FaaS environment. Contrasting existing solutions, we apply the model-free Q-Learning to determine required number of function instances for the workload demand that eventually reduces number of on-demand cold starts. Furthermore, the existing solutions takes advantage of either continuous pinging, pool-based approaches, container migration and network building or exploit platform-specific implementations like provisioned concurrency while failing to experiment with CPU-intensive real-world application workloads. Similar to \cite{agarwal2021reinforcement}, our work utilizes available resource-based metrics and response failure rate to accomplish the learning, but improves over the discussed approach. Contrasting to their model, we formulate the problem of cold starts as an optimisation approach to proactively spawn the required function instances and minimize frequent, on-demand cold starts. 

As part of their Q-learning model, the study used fixed-value constants in the reward modelling and we address this issue by carefully analysing the problem and curate it as a threshold-based reward system. Additionally, we experiment with \textit{matrix multiplication} function, that can be used as part of image processing pipeline, to train and evaluate our agent and utilise the open-sourced function invocation trace \cite{shahrad2020serverless} by Azure. Further we describe our design decisions and utilize constants based upon the trial-error analyses. The successful learning of the agent resulted in the preparation of near to optimal function instances in a timeframe to reduce the on-demand function creation or cold starts and improve the platform's throughput. A summary of related works is presented in Table \ref{relatedWork}.

\begin{landscape}
\begin{table}[!htbp]
\caption{Related work summary}
% \centering
\begin{tabular}
{|p{1cm}|p{3cm}|p{3cm}|p{3cm}|p{4cm}|p{3cm}|}
% {|l|l|l|l|l|l|}
\hline
\textbf{Work} & \textbf{Name} & \textbf{Platform} & \textbf{Solution Focus} & \textbf{Strategy} & \textbf{Application Type} \\ \hline
\cite{vahidinia2020cold} & - & AWS Lambda & Cold start latency & Optimising environments \& function pinging
& Concurrent \& sequential CPU \& I/O intensive \\ \hline
\cite{xu2019adaptive} & AWU \& ACPS & Kubernetes & Cold start latency \& resource wastage & Invocation prediction (LSTM) \& Container pool
 & Function chain model \\ \hline
\cite{lin2019mitigating} & - & Knative & Cold start frequency & Container pool \& pod migration & Single function model \\ \hline
\cite{bermbach2020using} & Naïve, Extended \& Global Approach & AWS Lambda, Apache Open Whisk & Cold start frequency & Orchestration middleware & Function chain model \\ \hline
\cite{mohan2019agile} & Pause Container Pool Manager & Apache OpenWhisk & Cold start latency& Container Pool & Function chain model \\ \hline
\cite{solaiman2020wlec} & WLEC & OpenLambda & Cold start latency & Container Pool & Single function model \\ \hline
\cite{mahajan2019exploiting} & - & AWS & Cold start latency & Container migration \& content similarity & Single function model \\ \hline
\cite{schuler2021ai} & - & Knative & Cold start frequency & AI-based container concurrency & Emulated CPU \& I/O intensive\\ \hline
\cite{silva2020prebaking} & Prebaking & OpenFaas & Cold start latency  & CRIU process snapshot & Single function model \\ \hline
\cite{vahidinia2022mitigating} & - & OpenWhisk & Cold start frequency  & RL-based idle window \& LSTM based container pre-warming & Single function model \\ \hline

\cite{benedetti2022reinforcement} & - & OpenFaas & Function Scaling  & RL \& SLA-based configuration & Single function model \\ \hline
Our work & - & Kubeless & Cold start frequency & AI-based function \& throughput metrics & Single function model\\ \hline
\end{tabular}
\label{relatedWork}
\end{table}
\end{landscape}

\section{System Model and Problem Formulation}
\label{section3}
FaaS is an event-driven cloud service model that allows stateless function deployment. It delivers high scalability and scale-to-zero feature being economical to infrequent demand patterns. New functions {$n_i$}, where {$1 \leq n_i \leq N$} and {$N$} is the maximum scale, are instantiated on-demand to serve the incoming load (scale up) and removed (scale down) when not in use after a certain time span or below a configured, resource-based threshold metric value for every {$i$} iteration window. The preparation time of function containers i.e., cold start {$C_t$}, adds to the execution time of a request. These frequent on-demand cold starts result in an increased computation pressure on existing resources, neglecting expected average CPU utilisation ({$\phi_o$}), and expected request failure rate ({$\tau_o$}). Therefore, an intelligent, learning-based solution is proposed to address them. 

\begin{figure}[!h]
         \centering
         \includegraphics[width=0.9\textwidth, height=9cm]{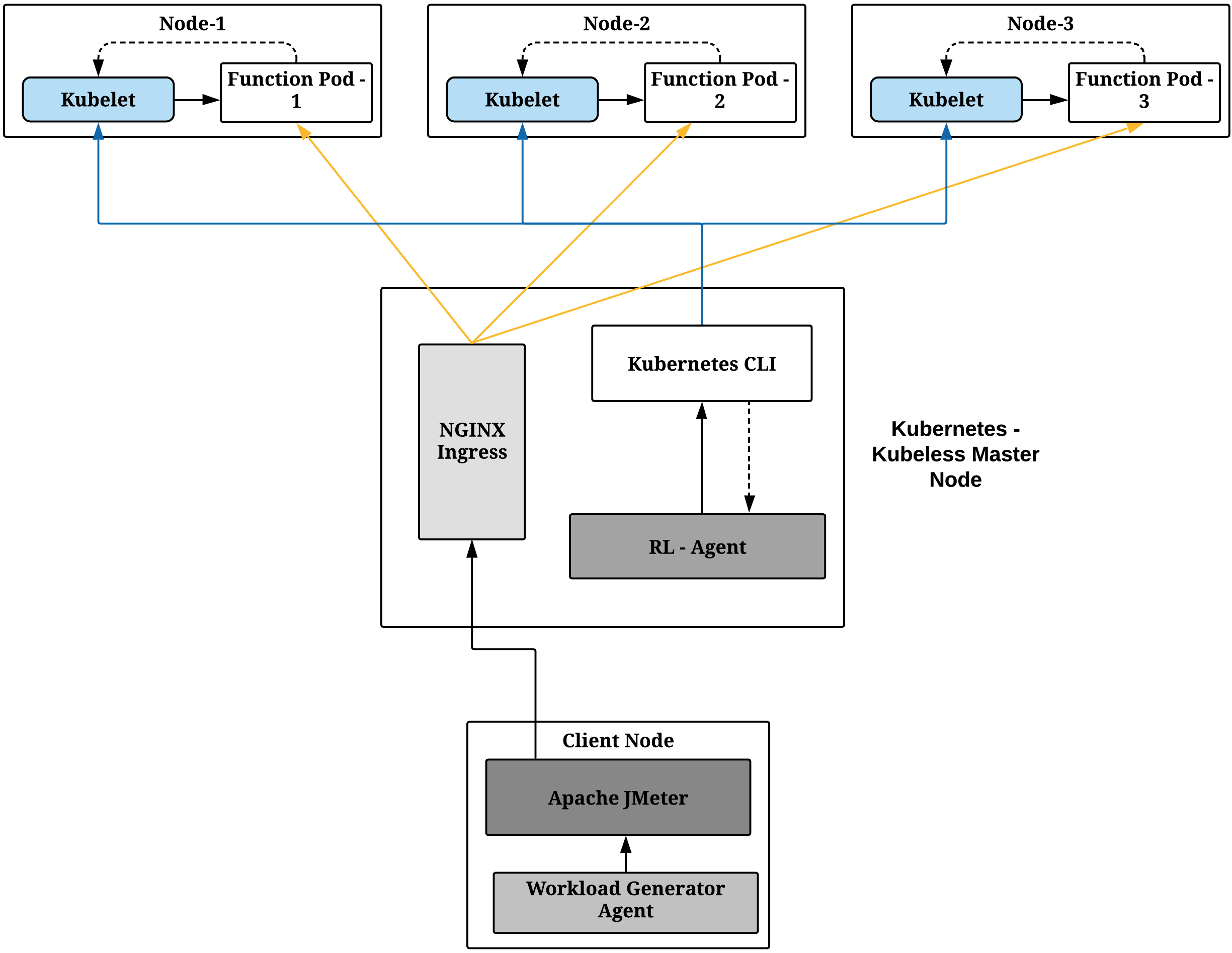}
             \caption{System Model}
        \label{systemModel}
    \end{figure}

\begin{figure}
         \centering
         \includegraphics[width=0.6\textwidth , height=7cm]{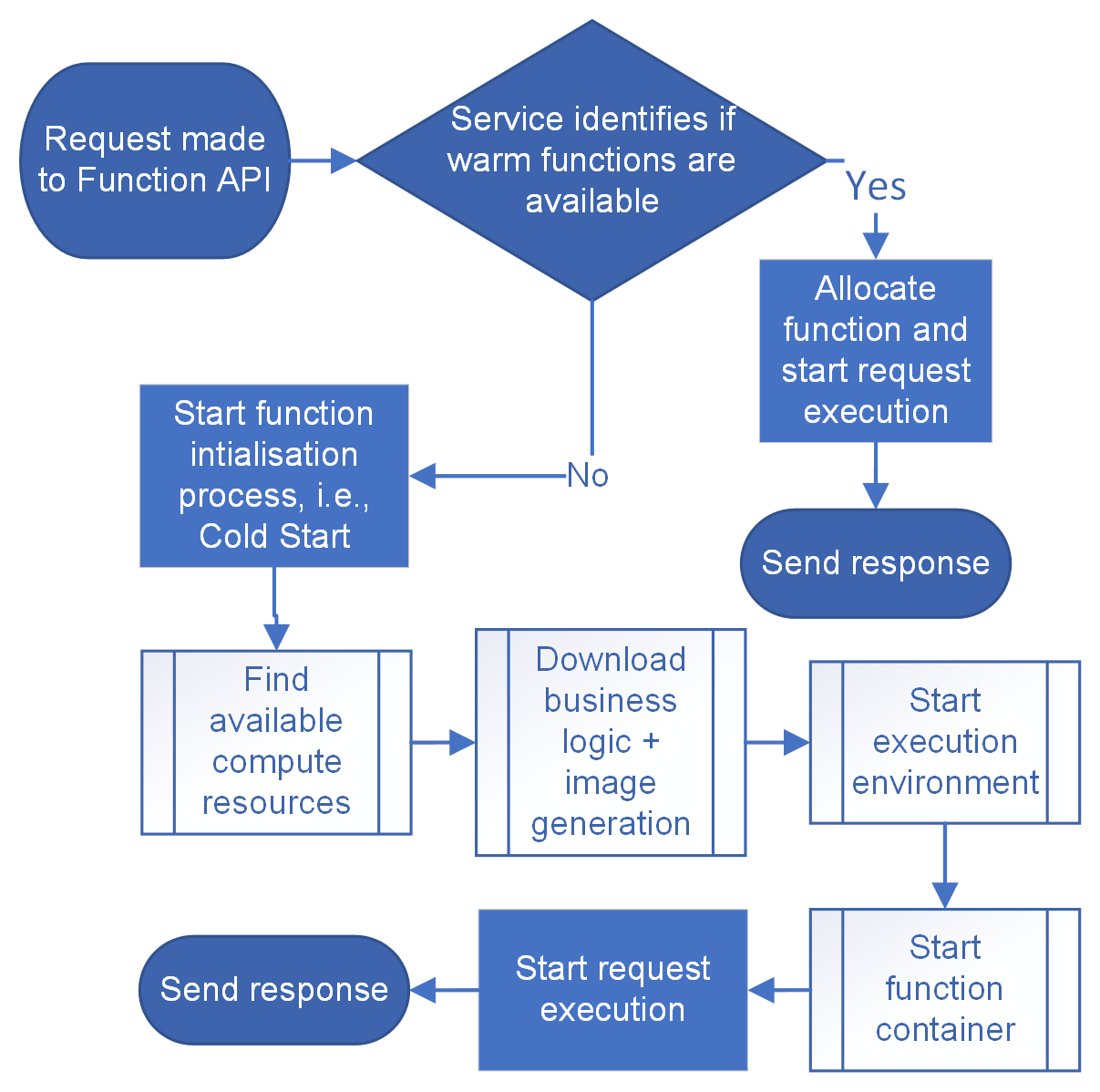}
             \caption{Function Warm Start \& Cold Start workflow}
        \label{csWorkflow}
     \end{figure}

In this study, we consider Kubeless, an open-source Kubernetes-native serverless platform that leverages Kubernetes primitives to provide serverless abstraction. It wraps function code inside a docker container with pre-defined resource requirements i.e. {$RR_f = (cpu_f, mem_f, tout_f)$} and schedules them on worker nodes. Similar to commercial FaaS providers, Kubeless has an idle-container window of 5 minutes to re-use functions and scales down to a minimum of one function if the collected metrics (default 15 seconds window) are below the set threshold. We take into account the general illustration of FaaS platform and consider a stochastic incoming request pattern {$D = \{d_1, d_2,\dots, d_i\}$} with {$d_i$} requests in {$i^{th}$} iteration window. We analyze the request pattern for a timeframe {$T$} divided in {$i$} iteration windows of duration {$t_i$}. The system model of the examined scenario is depicted in Fig. \ref{systemModel}. The workflow of the potential cold start is explained in Fig. \ref{csWorkflow}.

\subsection{Problem Formulation}
We formulate the function cold start as an optimization problem aimed at minimizing the number of cold starts (Eq. \ref{optimisation}) by preparing required instances, beforehand and aid the agent in learning a policy to reduce the request failure rate while maintaining average CPU utilisation. 
\begin{equation}
\label{optimisation}
\min_{\phi, \tau, d_i} \quad (n_i)
\end{equation}
\textbf{such that} \\
\begin{equation}
\tau_{d_i} < \tau_{o}; \
\phi_{d_i} < \phi_o  
\end{equation}

A cold start happens when there are no function instances available on the platform to deal with the incoming request and a new function instance is requested from the platform. FaaS services scale horizontally as per resource-based thresholds to be agile, usually considering the function's average CPU utilisation. Therefore, the goal of optimization is to assess the incoming request pattern {$d_i$} for an application task in {$i^{th}$} iteration window and configure a policy to prepare functions beforehand, considering actual and expected average CPU utilisation ({$\phi_{d_i} \& \phi_o$}) and request failure rate ({$\tau_{d_i} \& \tau_o$}). Since the preparation time, {$C_t$} remains similar for individual function containers, we focus on optimizing the frequency of cold start {$n_i$} for an individual iteration window.

With easy to implement and economical service model, enterprises are accommodating critical tasks like user verification, media processing, parallel scientific computations, anomaly detection and event-driven video streaming into the serverless paradigm. To assess the necessity of a dynamic solution, we consider \textit{matrix multiplication} as workload, which is an important task in image processing workflow.

\subsubsection{Reinforcement Learning model}
In a model-free Q-Learning process, the agent learns by exploring the environment and exploiting the acquired information. The core components of the environment are state, action, reward and agent. The environment state represents the current visibility of the agent and is defined as a Markov Decision Process (MDP) \cite{sutton1998reinforcement}, \cite{ashraf2018reinforcement} where future environment state is independent of past states, given the present state information. Actions are the possible set of operations that the agent can perform in a particular state. Additionally, rewards are the guiding signals that lead the agent towards the desired goal by performing actions and transitioning between environment states. The agent maintains a Q-value table to assess the quality of action through obtained reward for the respective state and utilize it for future learning.
Therefore, we propose a modelling scheme for the RL environment that is leveraged by a Q-Learning agent to learn a policy for function preparation.

We model the RL environment's state as $s_i= (\hat{n}_i,\phi_{d_i},\tau_{d_i})$ where {$\phi_{d_i}$} represents the average CPU utilisation of the available {$\hat{n}_i$} function instances, {$\tau_{d_i}$} represents the response failure rate, and {$i$} is the iteration window during a timeframe {$T$}. The agent's task is to prepare the estimated number of function instances in the upcoming iteration window either by exploring or exploiting the suitable actions. These actions of adding or adjusting the number of function instances, compensate for any expected cold starts from the incoming demand and help to improve the throughput of the system. Therefore, we define the agent's action as the number of function instances, {$n_i$}, to be added or removed from currently available functions {$\hat{n}_{i-1}$} and represent it as a set $a_i =$ $\{n_i | 1 \leq (\hat{n}_{i-1} + a_i) \leq N\}$. This heuristic helps the agent to control the degree of exploration by maintaining the number of functions within the threshold $N$, that is adapted based on deployed infrastructure capacity. Hence, we map the function resources and relevant metrics to RL environment primitives.

The motive of the RL-based agent is to learn an optimal policy, and we structure the rewards over resource-based metrics {$\phi_{d_i}$}, function response failure rate {$\tau_{d_i}$}, and expected threshold values {$(\phi_o$ and $\tau_o)$}. It evaluates the quality of action {$a_i$} in state {$s_i$} by keeping a value-based table, i.e., Q-table, that captures this information for every {$(s_i,a_i)$} pair. After executing the action, the agent waits for the duration of the iteration window and receives a delayed reward {$r_i$}, expressed based on the difference between the expected and actual utilisation and failure rate values, as shown in Eq. \ref{reward}.

\begin{equation}
\label{reward}
        r_i = \dfrac{(\phi_o - \phi_{d_i}) + (\tau_o - \tau_{d_i})}{\hat{n}_i}
\end{equation}
and the Q-table is represented as a matrix (Eq. \ref{matrix}) of dimension {$S \times A$}.

\begin{equation}
\label{matrix}
Q_{(S_n \times A_m)} = 
 \begin{bmatrix}
s_1,a_1 & \dots & s_1,a_m\\
\vdots & \ddots & \vdots\\
s_n,a_1 & \dots & s_n,a_m
\end{bmatrix}   
\end{equation}

\section{Q-Learning for Cold Start Reduction} 
\label{section4}
In this work, we apply model-free Q-learning algorithm in FaaS paradigm to reduce frequent on-demand function cold starts. We select this algorithm due to its simple and easy implementation, model interpretability, strong theoretical convergence guarantees, ability to process the perceived information quickly using the Bellman equation and its adaptability to other advanced algorithms like Deep Q-Learning (DQN). As discussed in Section \ref{section3}, we model the process of creating required function instances as an MDP and map the serverless computing primitives to RL agent's environment, state and actions. We explore and exploit an off-policy RL algorithm to reduce the on-demand function cold starts and determine the required function instances with the intuition of it being easy, simple to implement, less complex with stable convergence in a discrete action space. The proposed approach has two phases: an agent training phase and a testing phase. \emph{Algorithm \ref{qlearnAlgo}} demonstrates the agent training workflow. The environment setup process precedes the agent training, where the agent interacts with the environment and obtains information. After initial setup, the agent is trained for multiple epochs or timeframes where it assesses the function demand {$d_i$} over individual iteration windows {$i$} and ascertains appropriate function instances. During an iteration window {$i$}, the agent observes the environment state {$s_i$}, selects an action {$a_i$} according to {$\epsilon$}-greedy policy. This greedy policy helps the agent to control its exploration and selects a random action with {$\epsilon$} probability, otherwise exploiting the obtained information. This exploration rate is a dynamic value and decays with ongoing learning to prioritise the acquired information.

\begin{algorithm}
 \caption{Q-Learning for Cold Start Reduction}   
 \label{qlearnAlgo}
\begin{algorithmic}
\Require Initialise Environment variables 
\Ensure Initialise Q-Table, $decayRate$, $\epsilon$, $epoch$
% \State $decayRate = 0.0025$
\State $\epsilon = 0.01 + 0.99\mathrm{e}^{(-decayRate \times epoch)}$
\State Repeat for each $Training Epoch$
\State $epoch \gets epoch + 1$
% \State $t^l \gets T_D/l $
\While{$t_{elapsed} < T$}
        \State $s_i \gets currentState(\hat{n}_i,\phi_{d_i},\tau_{d_{i}},i)$ 
        \State $a_i \gets$ choose using $\epsilon$-greedy policy from Q-Table
        \State Scale \& wait for $i^{th}$ iteration window
        \State $r_i \gets calculateReward(\phi_{d_i},\tau_{d_i})$
        \State $s_{i+1} \gets getNewState(\hat{n}_{i+1},\phi_{d_{i+1}},\tau_{d_{i+1}},{i+1})$
        \State $Q(s_i,a_i)\gets(1-\alpha)Q(s_i,a_i +\alpha(r_i+\gamma \max_a Q(s_{i+1},a_{i}))$
        \State $t_{elapsed} = t_{elapsed} + t_i$
    \EndWhile
\end{algorithmic}
\end{algorithm}

After performing the selected action, the agent waits for duration {$t_i$} of an iteration window to obtain the delayed reward {$r_i$}, calculated using the relevant resource-based metrics {$\phi_{d_i}$} and function failure rate {$\tau_{d_i}$}. This reward helps the agent in action quality assessment, and it combines the acquired knowledge over previous iterations using the Bellman Equation (Eq. \ref{bellman}). It is the core component in learning as it aids Q-value or Q-table updates and improves the agent's value-based decision-making capability. The equation uses two hyper-parameters learning rate, {$\alpha$} and discount factor, {$\gamma$}. The learning rate signifies the speed of learning and accumulating new information, and the discount factor balances the importance of immediate and future rewards.  

\begin{equation}
\label{bellman}
        Q(s_i,a_i) = (1 - \alpha)Q(s_i,a_i) + \alpha(r_i + \gamma \max_a Q(s_{i+1},a_{i}))
\end{equation}

The agent then evaluates and adjusts the Q-value in Q-Table based upon the delayed reward for the corresponding ({$s_i,a_i$}) pair. The agent continues to analyse the demand over multiple iteration windows, selecting and performing actions, evaluating delayed rewards, assessing the quality of action and accumulating the information in Q-table, and repeating this process over multiple epochs for learning. Once the agent is trained for sufficient epochs and the exploration rate has decayed significantly, we can exploit obtained knowledge in the testing phase.

In the testing phase, the agent is evaluated using a demand pattern for the \textit{matrix multiplication} function and the Q-table values guide the agent in taking informed actions. The agent determines the current environment state and obtains the best possible action i.e. action with the highest Q-value for the corresponding state, and adjusts the required number of functions based on its understanding of the demand. We hypothesise that there exists a relationship between throughput and function availability to serve incoming requests. Therefore, we evaluate the agent's performance by considering metrics such as system throughput, function resource utilisation and available function instances. We further hypothesise that the RL-based agent learns to prepare and adjust required number of functions beforehand and improve the throughput while keeping the function's resource utilisation below the expected threshold.

\begin{table}[!th]
\caption{System Setup Parameter values}
\centering
\begin{tabular}{|c|c|}
\hline
\textbf{Parameter Name} & \textit{\textbf{Value}}     \\ \hline
Kubernetes version      & v1.18.6                     \\ \hline
Kubeless version        & v1.0.6                      \\ \hline
Nodes                   & 4                           \\ \hline
OS                      & Ubuntu 18.04 LTS            \\ \hline
vCPU                    & 4                           \\ \hline
RAM                     & 16 GB                       \\ \hline
Workload                & Matrix Multiplication ({$m \times m$}) \\ \hline
m                       & 1024                        \\ \hline
\end{tabular}
\label{systemParam}
\end{table}

\section{Performance Evaluation}
\label{section5}
In this section, we provide the experimental setup and parameters, and perform an analysis of our agent compared to other complementary solutions. 

\subsection{System Setup}
We set up our experimental test-bed as discussed in Section \ref{section3}, using NeCTAR (Australian National Research Cloud Infrastructure) services on the Melbourne Research Cloud. We configure Kubernetes (v1.18.6) and Kubeless (v1.0.6) on a service cluster of 4 nodes, each with Ubuntu (18.04 LTS) OS image, 4 vCPUs, 16 GB RAM, and 30 GB of disk storage to perform the relevant experiments. Typical serverless applications expect high scalability for their changing demands and can be compute-intensive, demanding a considerable amount of resources such as CPU, memory, or time to execute. These factors add to frequent cold starts on the platform by keeping the available functions or resources busy while requesting new functions for the subsequent workload demand. We use Python-based matrix multiplication (1024 pixels x 1024 pixels) to mimic the image processing task as our latency-critical application to deploy serverless functions. 

The experimental setup mimics real-time application demand experienced in commercial FaaS platforms \cite{galstyan2004resource}, \cite{shahrad2020serverless}. We consider a single function invocation trace from the open-source Azure function data \cite{shahrad2020serverless} and downsize it according to our resource capacity. We deploy the Apache JMeter load testing tool to generate the HTTP-based requests and randomize its request ramp-up period to guarantee the changing demand pattern for our workload. Also, we collect the relevant resource-based metrics and throughput information via Kubernetes APIs. Table \ref{systemParam} summarises the parameters used for system set-up.

% \begin{figure}[!htb]
% \centering
    \begin{figure}[htb]
    \centering
        \includegraphics[width=0.6\textwidth, height=4cm]{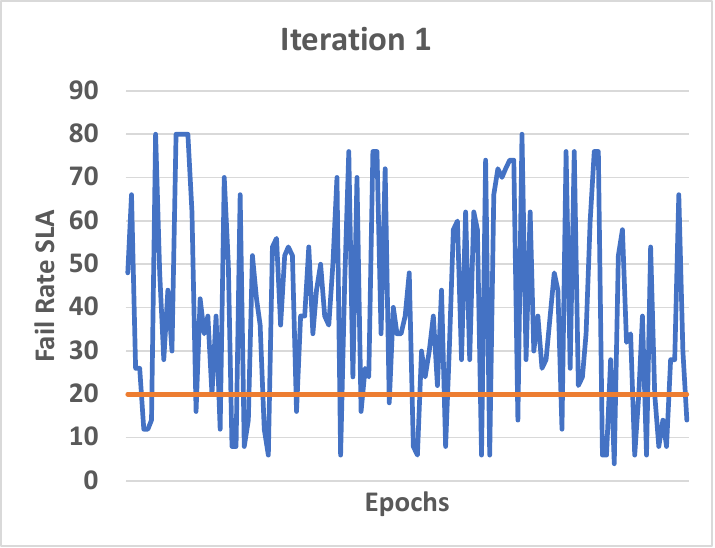}
        \caption{Training iteration 1}
        \label{training1}
    \end{figure}
    % \hfill
        \begin{figure}
        \centering
        \includegraphics[width=0.6\textwidth, height=4cm]{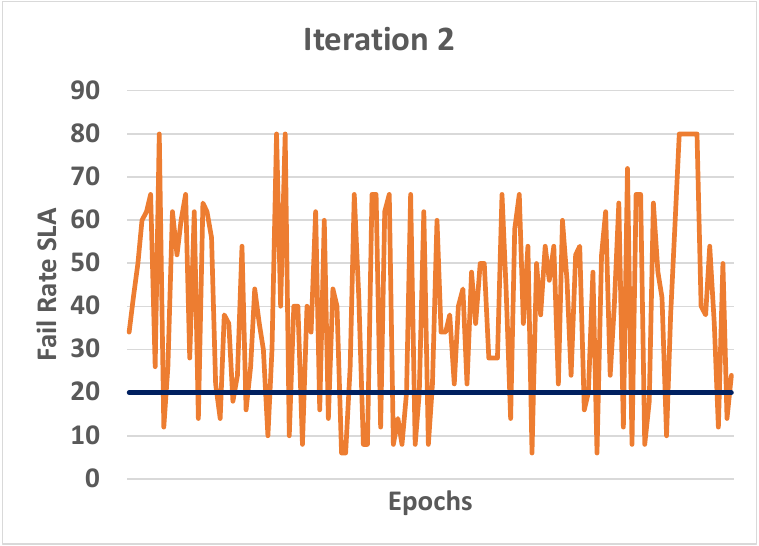}
        \caption{Training iteration 2}
        \label{training2}
    \end{figure}
    % \hfill
    \begin{figure}[htb]
        \centering
        \includegraphics[width=0.6\textwidth, height=4cm]{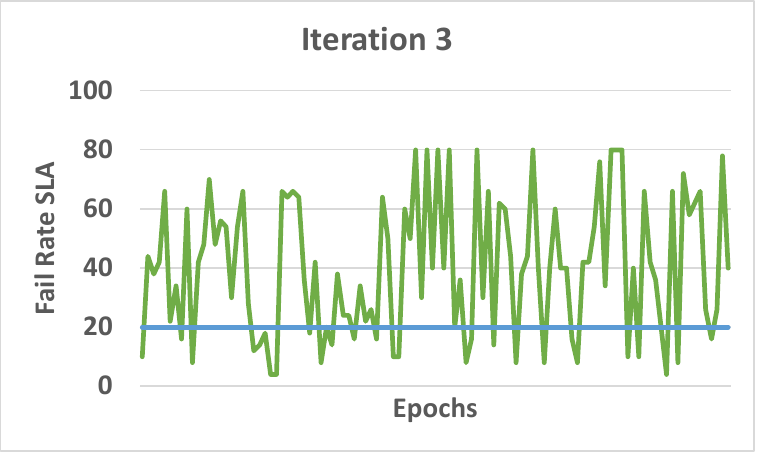}
        \caption{Training iteration 3}
        \label{training3}
    \end{figure}
    % \hfill
        \begin{figure}[htb]
        \centering
        \includegraphics[width=0.6\textwidth, height=4cm]{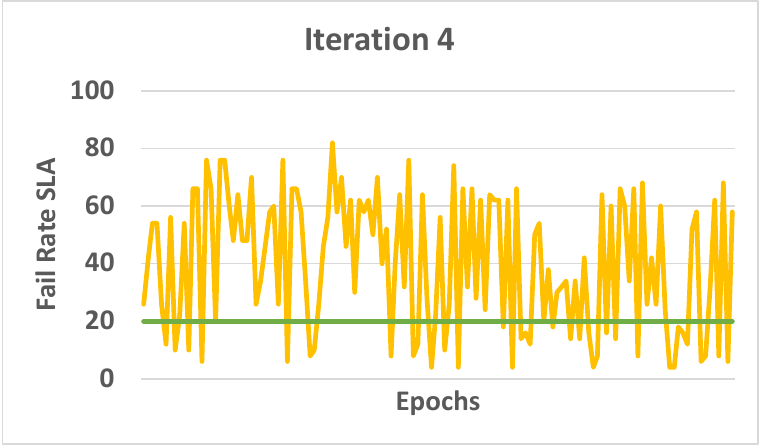}
        \caption{Training iteration 4}
        \label{training4}
    \end{figure}
    
    \begin{figure}[htb]
    \centering
        \includegraphics[width=0.6\textwidth, height=4cm]{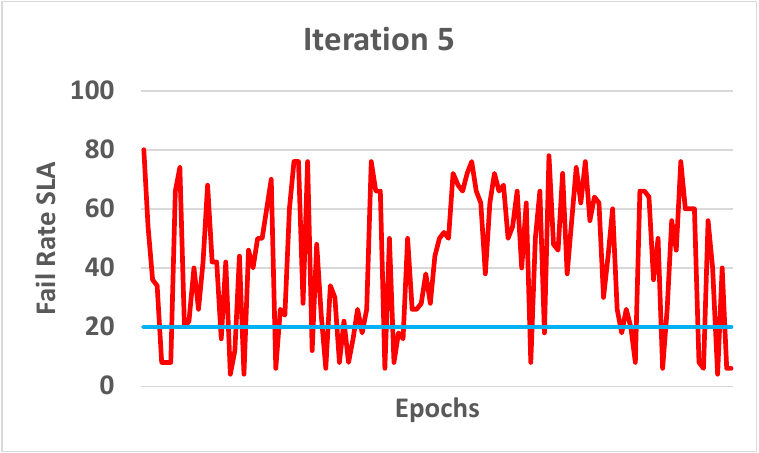}
        \caption{Training iteration 5}
        \label{training5}
    \end{figure}

\subsection{RL Environment Setup}
To initialize the proposed RL-based environment, we first analyze and set up the function requirements according to deployed resource limits. After preliminary analysis, we configure the function requirements as 1 vCPU, 128 MB memory, and 60 seconds function timeout, where timeout represents the maximum execution period for a function until failure. To experiment we assume a timeframe of 10 minutes to analyse the demand pattern of 100 requests during 5 iteration windows of 2 minutes. Based on the resource analysis and underlying Kubernetes assets we assume the function limit {$N=7$}. These constraints allow us to put a considerable load or pressure on the different techniques discussed and effectively evaluate them against each other. 

As discussed in Section \ref{section3}, the RL-environment components depend upon resource metrics (average CPU utilisation), response failure rate, number of available functions and expected threshold values, summarized in Table \ref{rlparam}. Since the proposed agent maintains a Q-table, these considerations help to minimise the risk of state-space explosion related to Q-Learning. The actions signify the addition or removal of functions based upon the function limit and the reward is modelled around the expected threshold values. We configure the Bellman Equation hyper-parameters: learning rate and discount factor as 0.9 and 0.99, respectively, based on the results of hyper-parameter tuning in \cite{agarwal2021reinforcement}. The agent is structured to explore the environment and exploit the acquired knowledge. We use {$\epsilon$}-greedy action selection policy to randomly select an action with initial {$\epsilon=1$} probability and exploit this information with a decay rate of 0.0025. These RL system parameter values were chosen after careful consideration of discussed workload and invocation pattern, according to the underlying resource capacity, and to showcase the applicability of RL-based agent in a serverless environment.

\begin{table}[!thb]
\caption{RL-Environment parameter values}
\centering

\begin{tabular}{|c|c|}
\hline
\textbf{Parameter} & \textit{\textbf{Value}} \\ \hline
$cpu_f,mem_f,tout_f$     & 1, 128M, 60 seconds  \\ \hline
$N$                & 7                       \\ \hline
$T$               & 10 minutes              \\ \hline
$i$                & 5                       \\ \hline
$t_i$                & 2 minutes                       \\ \hline
$\phi_o$                & 75\%                    \\ \hline
$\tau_o$                & 20\%                    \\ \hline
$\alpha$              & 0.9                     \\ \hline
$\gamma$              & 0.99                    \\ \hline
$\epsilon$   & 1                       \\ \hline
$decayRate$         & 0.0025                  \\ \hline
\end{tabular}
\label{rlparam}
\end{table}

\subsection{Q-Learning Agent Evaluation}

We train the RL-based agent for a timeframe of 10 minutes over 500 epochs to analyze an application demand and learn the ideal number of functions to reduce frequent cold starts. The agent is structured according to the RL-based environment design explained in section \ref{section3} and around the implementation constraints. The quality of the RL-based agent is evaluated during a 2-hour period to reduce the effect of any bias and performance bottlenecks. 

We assess the effectiveness of our approach against the default scaling policy and commercially used function keep-alive policy on the serverless platform. Kubeless leverages the default resource-based scaling (HPA) implemented as a control loop that checks for the specific target metrics to adjust the function replicas. HPA has a default query period of 15 seconds to check and control the deployment based on the target metrics like average CPU utilization. Therefore, the HPA controller fetches the specific metrics from the underlying API and calculates the average metric values for the available function instances. The controller adjusts the desired number of instances based on threshold violation but is unaware of the demand and only scales after a 15-second metric collection window. The expected threshold for function average CPU utilisation is set to be 75\% with maximum scaling up to 7 instances. Therefore, whenever the average CPU utilisation of the function violates the threshold, new function instances are provisioned in real-time, representing a potential cold start in the system. 

Also, HPA has a 5-minute down-scaling window and during that period resources are bound to the platform irrespective of incoming demand which represents potential resource wastage. Therefore, it is worthwhile to analyse the performance of the RL-based agent against the function queuing or keep-alive approach that keeps enough resources bound to itself for an idle-container window. 

\begin{figure}
    \centering
        \includegraphics[width=0.6\textwidth, height=4cm]
        {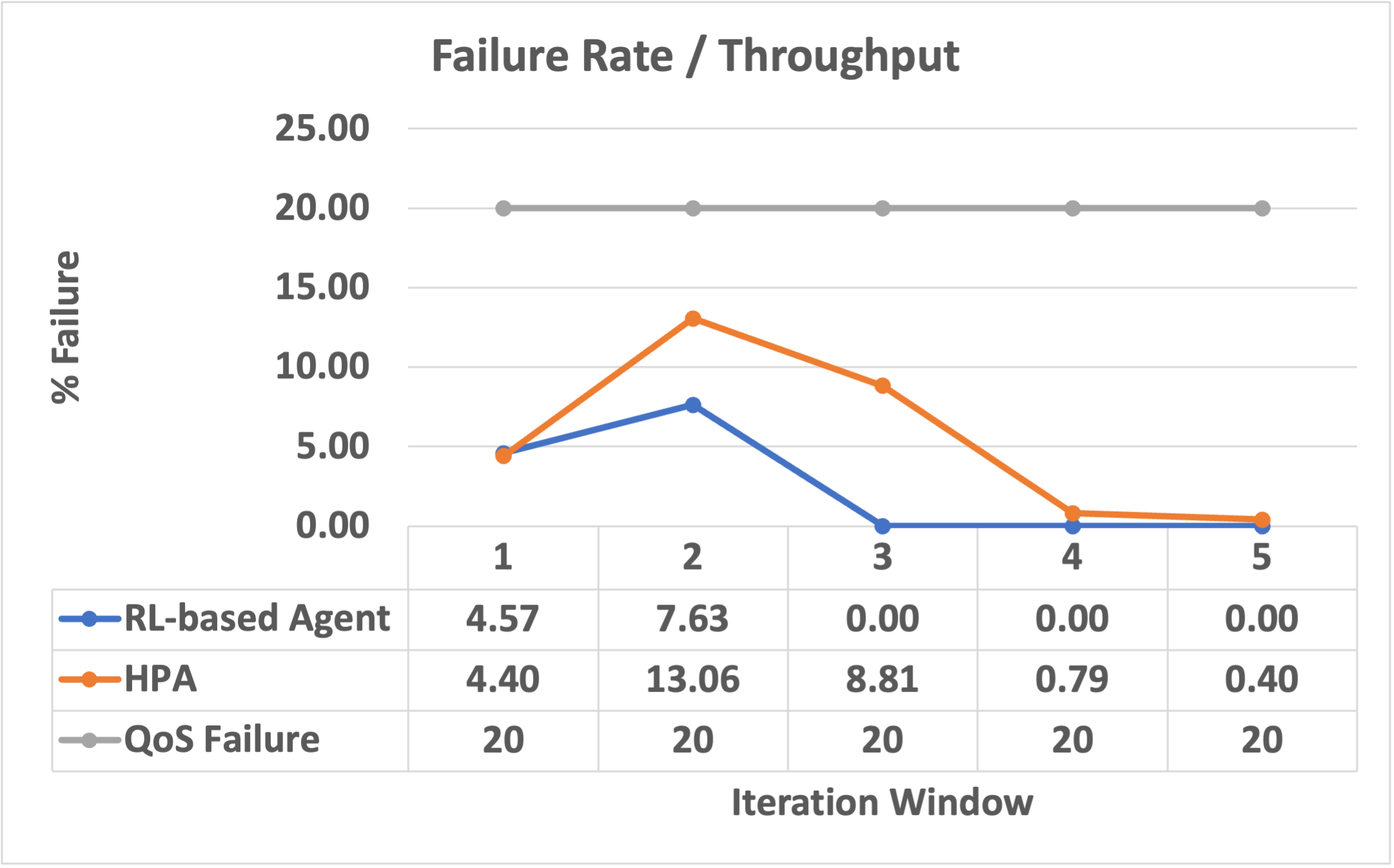}
        \caption{RL Agent v/s HPA: Failure Rate}
        \label{hpaQoS}
    \end{figure}

\begin{figure}
        \centering
        \includegraphics[width=0.6\textwidth, height=4cm]{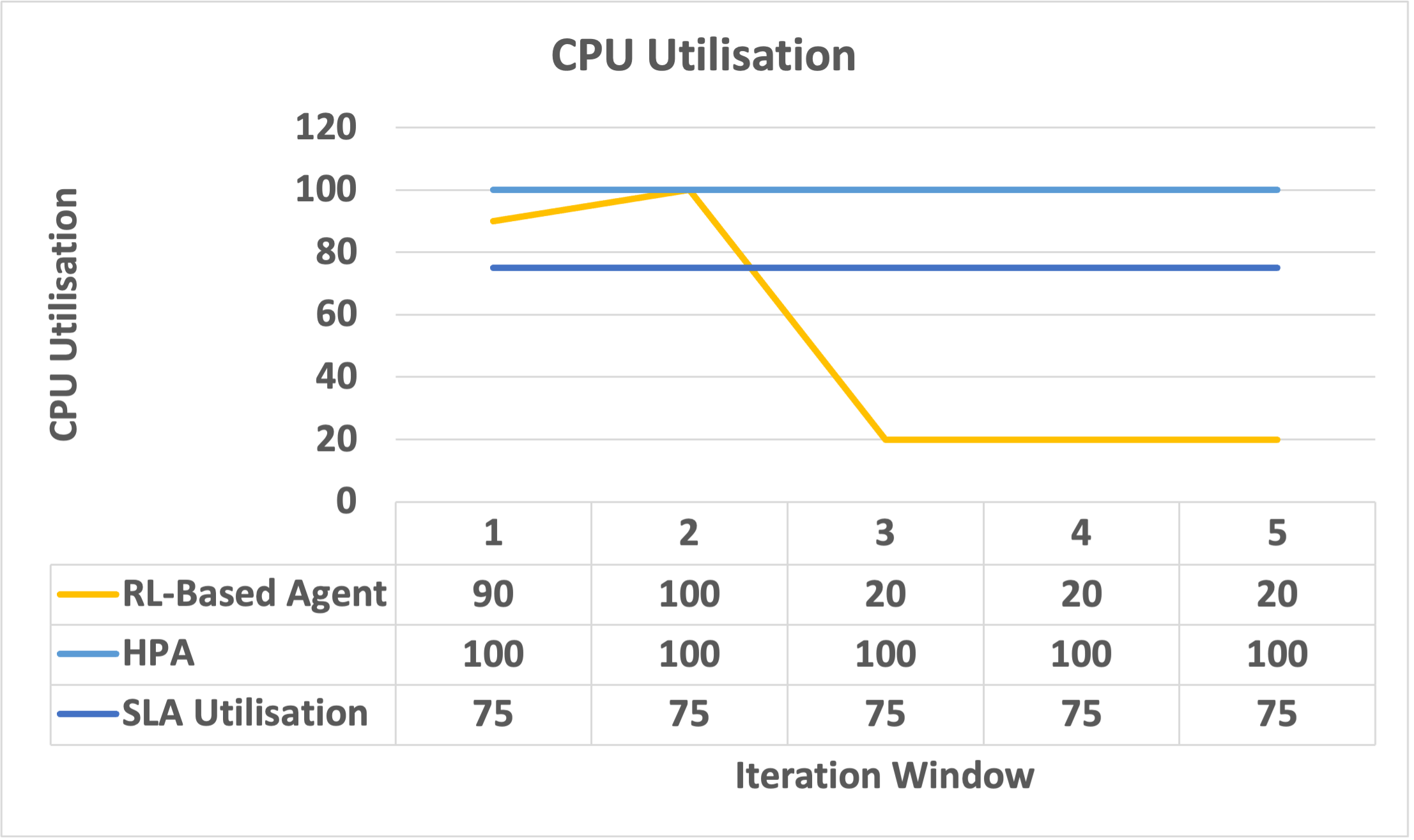}
        \caption{RL Agent v/s HPA: CPU Utilisation}
        \label{hpaSLA}
    \end{figure}

\begin{figure}
        \centering
        \includegraphics[width=0.6\textwidth, height=4cm]{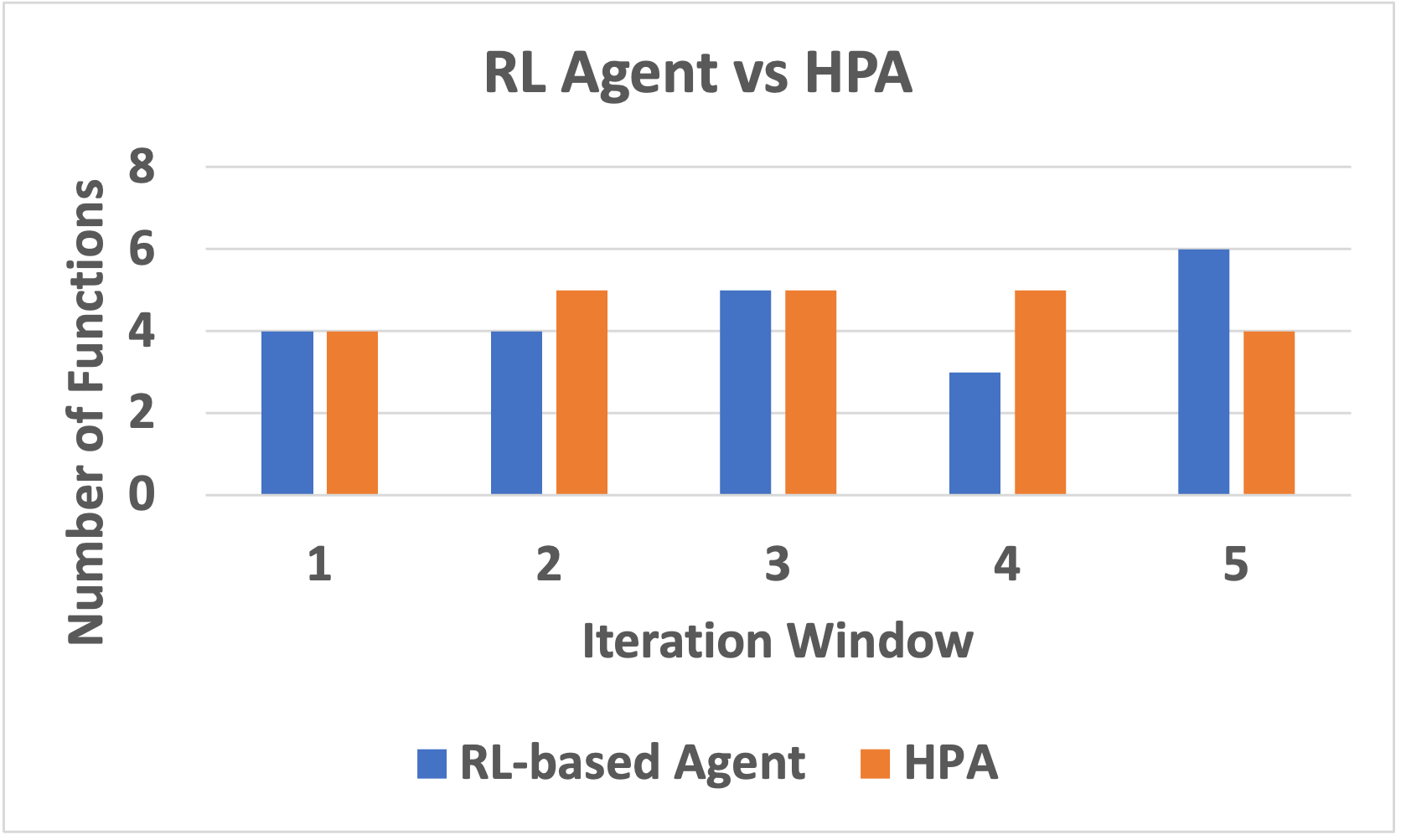}
        \caption{HPA: Function Provision}
        \label{hpaProvision}
    \end{figure}

Fig. \ref{training1},\ref{training2},\ref{training3},\ref{training4},\ref{training5} illustrates the learning curve of the agent over multiple epochs and we observe that the agent continuously attempts to meet the expected thresholds. This highlights the agent's capability to obtain positive rewards and move towards the desired configuration. We compare the RL-based agent with HPA and successfully demonstrate the agent's ability to improve the function throughput i.e., reduce the failure rate by up to 8.81\%, Fig. \ref{hpaQoS}. The RL-based agent further targets to maintain the expected CPU utilisation thresholds, Fig. \ref{hpaSLA}, by reducing CPU stress up to 55\% while determining the required function instances in Fig. \ref{hpaProvision}. For example, in Fig. \ref{hpaProvision} during iteration windows 1 and 2, the HPA scales functions based on CPU utilisation threshold, unaware of the actual requirement for upcoming iteration and results in resource wastage. Similarly, Fig. \ref{hpaProvision} illustrates the resource wastage by HPA during iterations 3 and 4.

\begin{figure}
        \centering
        \includegraphics[width=0.6\textwidth, height=4cm]{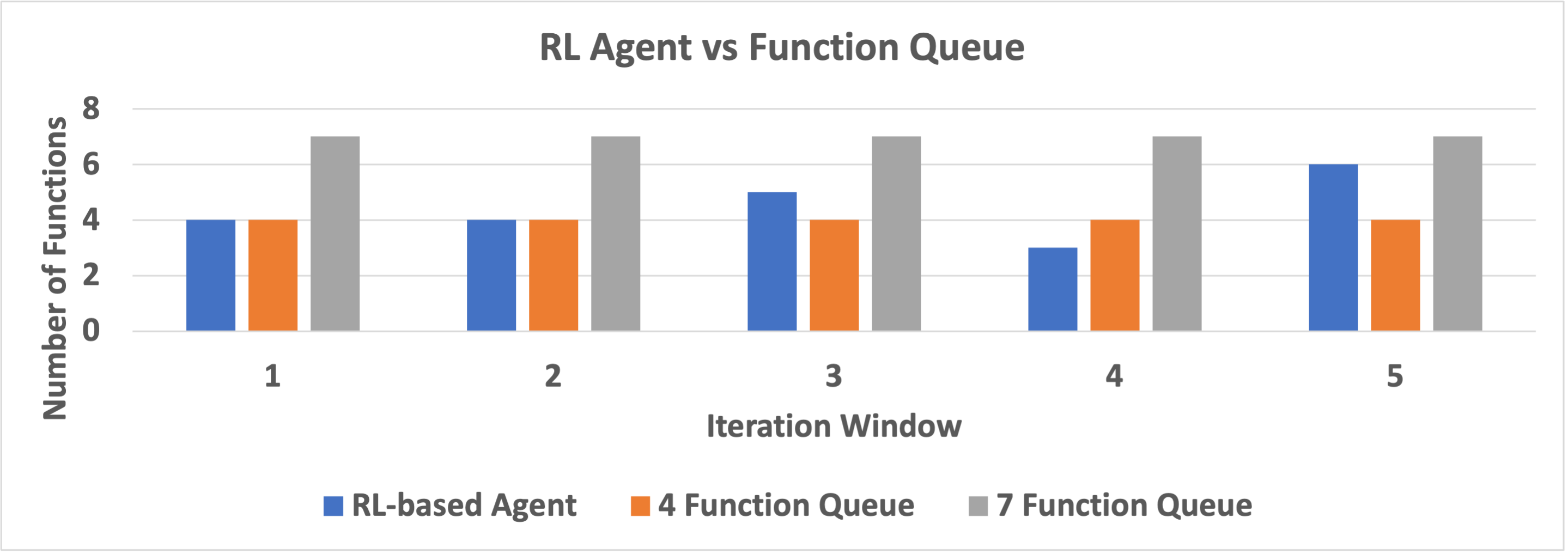}
        \caption{Function Queue: Function Provision}
        \label{queuePods}
    \end{figure}

\begin{figure}
    \centering
        \includegraphics[width=0.6\textwidth, height=4cm]{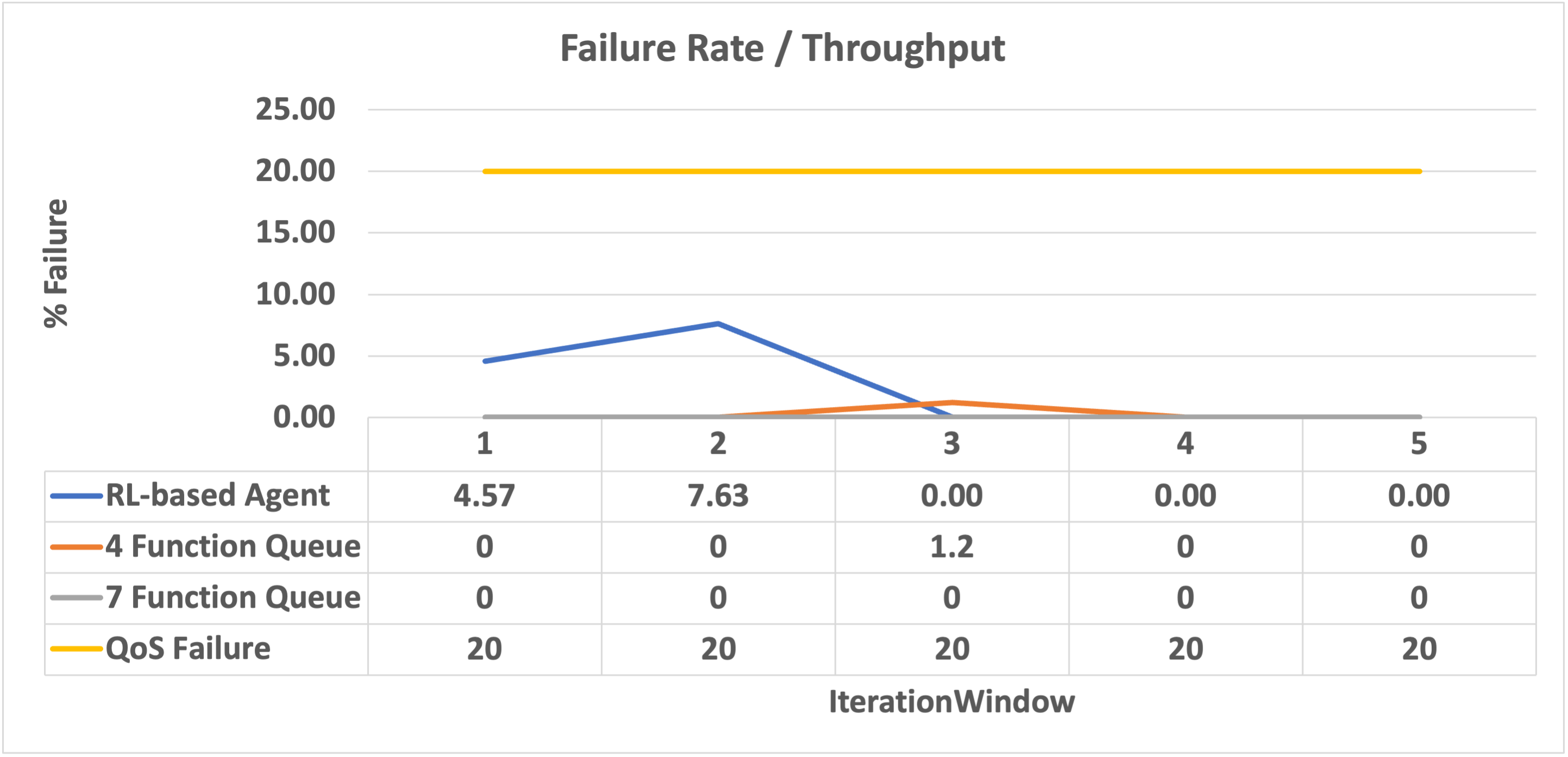}
        \caption{RL Agent v/s Function Queue: Failure Rate}
        \label{queueQoS}
    \end{figure}

\begin{figure}
    \centering
        \includegraphics[width=0.6\textwidth, height=4cm]{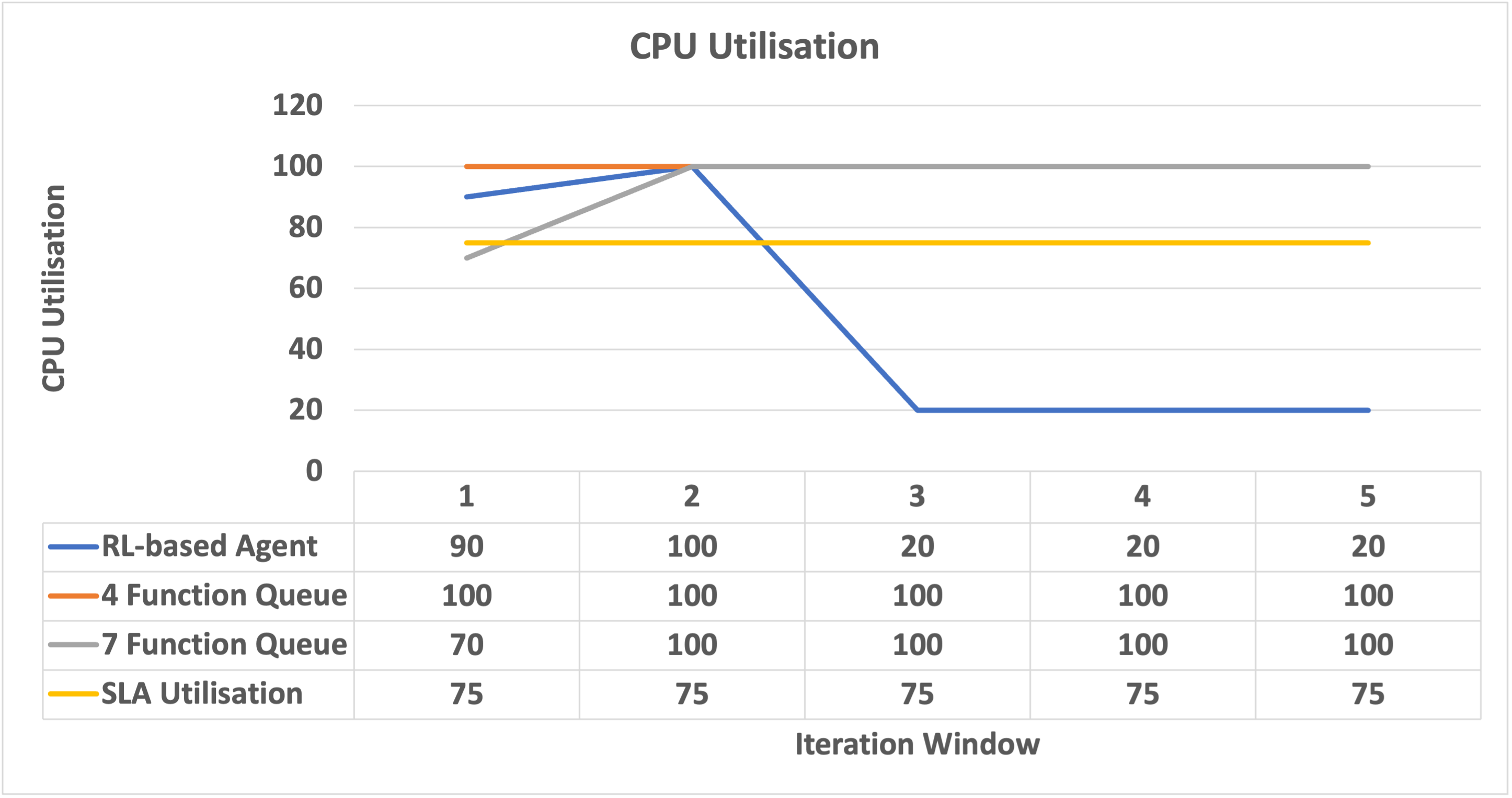}
        \caption{RL Agent v/s Function Queue: CPU Utilisation}
        \label{queueSLA}
    \end{figure}

Similar results are observed against function queuing or keep-alive policy, where we evaluate two queues with {$N = 4$ and $N = 7$}. The RL-based agent scales and prepares the function according to demand needs while the queue results in resource wastage of up to 37\%, as shown in  Fig. \ref{queuePods}. Although the queuing policy manages to reduce the request failure rate to zero, it is due to extra resources available, as depicted in Fig. \ref{queueQoS}, but can not be precisely captured by HPA metrics and shows over CPU utilisation of up to 50\% in Fig. \ref{queueSLA}. The proposed agent analysed the demand pattern by consuming sufficient function resources, preparing the ideal number of functions and trying to keep the desired CPU utilisation under control. Hence, the learning and testing analysis support our hypothesis that reducing on-demand cold starts can be directly linked to the throughput improvement.

% \begin{figure}
% \centering
    % \hfill
    % \hfill
    
    % \hfill
    
    % \hfill

\section{Discussion}
\label{section6}
Function cold start is an inherent shortcoming of the serverless execution model. Thus, we have proposed an RL-based technique to investigate the demand pattern of the application and attempt to reduce the frequency of function cold starts. The proposed agent performs better than the baseline approaches under a controlled experimental environment. But there are certain points to recollect associated with the real-time appropriateness of the proposed solution.

We leverage the RL environment modelling, specifically Q-Learning constraints \cite{sutton1998reinforcement}, \cite{zychlinski2019qrash}, and in general, these algorithms are expensive in terms of data and time. The agent interacts with the modelled environment to acquire relevant information over multiple epochs that signify a higher degree of exploration. Hence, as evidenced in the proposed work, for an RL-based agent to outperform a baseline technique, a training period of 500 epochs is exploited for satisfactorily analyzing the workload demand for a timeframe (10 minutes). Therefore, RL-based approaches are considerably expensive in practical applications with stringent optimization requirements.

A classical Q-Learning approach is applied to discrete environment variables \cite{sutton1998reinforcement}. To constrain the serverless environment within the requirements of the Q-Learning algorithm, we consider the discrete variables to model cold starts. The size of the Q-table is large and is a function of state space and action space. But with the expansion of state space or the action space, the size of the Q-table grows exponentially \cite{sutton1998reinforcement}, \cite{zychlinski2019qrash}. Therefore, Q-Learning experiences state explosion, making it infeasible to perform updates on Q-values and degraded space and time complexity.

The proposed agent analyzes individual application demand, so the learning can’t be generalized for other demand patterns and requires respective training to be commissioned. Furthermore, the agent is trained for 500 iterations and evaluated, but the chance of exploring every state is bleak with limited iterations of training. Therefore, the agent expects to be guided by certain approximations to avoid acting randomly. The agent utilizes resource-based metrics that affect the cold starts, so the availability of relevant tools and techniques to collect instantaneous metrics is essential. Also, the respective platform implementation of a serverless environment, such as metrics collection frequency, function concurrency policy, and request queuing, can extend support to the analyses.

The difference between the approaches can be attributed to the following characteristics of the proposed RL-based agent – 
\begin{enumerate}
    \item The process of elimination of invalid states during the RL environment setup and lazy loading of Python, helps the agent to productively use the acquired information about the environment. 
    \item Although the RL-based agent outperforms HPA and function queue policy, there is a lack of function container concurrency policy. The CPU-intensive function workload is configured with an execution time of 60 seconds and thus affected by the concurrency control of the instance.
    \item The composition of state space and reward function incorporates the effect of failures during the training, and therefore, the agent tries to compensate for the failures in consequent steps of learning by exploiting the acquired knowledge. 
\end{enumerate}
On the account of the performance evaluation results, we can adequately conclude that the proposed agent successfully outperforms competing policies for the given workload and experiment settings. We strengthen this claim by analysing the training and testing outcomes of the RL-based agent, focused on examining the workload pattern to reduce request failure which is a direct consequence of appropriate function instances representing reduced function cold starts.

\section{Future Research Directions}
\label{section7}

As part of the future research in broader serverless computing domain, various aspects of resource management needs to be addressed. These could be broadly categorised under workload estimation and characterisation, resource scheduling, and resource scaling. Following the increased adoption of FaaS, challenges such as function cold start delay, co-located function interference, lack of QoS guarantees, user pricing model, runtime limitations and support for specialised hardware, resource efficiency and workload management emerge as concerning factors for the success of serverless computing \cite{mampage2022holistic}. With the changing dynamics of application workload, its resource demand and introduction of AI/ML models in real-time systems, there is need for adaptive and proactive methods to tackle inefficient resource management in serverless computing. Additionally, there is a need for autonomous resource management system to completely offload the function scheduling, resource allocation and resource scaling tasks to offer a true serverless experience for developers and utilise the provider resources to the maximum. However, a guarantee of QoS objectives is still warranted in terms of service latency, throughput, fault tolerance and cost incurred by the user. Other aspects of FaaS that require meticulous thinking with adoption in edge and fog computing environments are data privacy and data locality concerns that further reduces the flexibility of FaaS application model. 

Specifically, to address the issue of function cold start delays, an impact of other important aspects such as function memory allocation, language runtime, deployment size, programming convention and function characteristics in conjunction with different techniques can be explored as discussed below:

\begin{itemize}
    \item In the industrial works like \cite{powertuning1}, \cite{powertuning2}, an insight is provided for the serverless AWS Lambda functions and how an improvement can be made to its cold start latencies and operating costs with careful function initialisation phase consideration. 
    \item An in-depth analysis of function fusion or monolithic function development on resource requirements and cold start of function workflows can also be explored with respect to the performance and run-time costs of the function. 
    \item Furthermore, a trade-off analysis of techniques like function pre-baking, provisioned concurrency and reserved concurrency can also be valuable in performance optimisation of serverless function and associated application workflows. 
    \item In addition to CPU-intensive functions, impact of cold start on different classes of functions like memory-intensive, I/O-intensive and network-intensive functions, utilised for application domains such as AI/ML model training and inference, and media processing, can also be explored. 
    \item Consequently, a cost and performance analysis of RL-based agents can also be made for the function cold start reduction in these application cases. Similar to Q-Learning, the application of other policy-based techniques such as SARSA, which is known to converge faster than Q-Learning, can also be experimented with in the domain of the cold start problem. As an adaptation of Q-Learning, the proposed solution includes discrete values over continuous values for state representation. In this context, to avoid the problem of state space explosion, function approximation techniques such as DQNs, Proximal Policy Optimization (PPO) and other complex deep learning methods such as Soft-Actor-Critic (SAC) or Recurrent models \cite{Drescle} can also be leveraged to estimate the information about optimal actions. 
    \item To further explore the evolving technological space, an integration of Generative Adversarial Networks for synthetic training data generation or Large-Language-Models can be explored for pro-active cold start reduction using the historical data.
\end{itemize}

\section{Summary and Conclusions}
\label{section8}
FaaS model executes the piece of code inside a container, known as a function and prepares new function containers on demand. New function containers undergo an initialisation process that puts together all the essential components before executing the function handler. This bootstrapping process consumes time in the order of a few seconds, known as function cold start and introduces a delay in the response of the function container. 

This work visits the problem of function cold start by addressing frequent cold starts and analysing the application demand through an RL technique. We leverage the services of Apache JMeter to produce varying incoming request patterns and a CPU-intensive function workload to complement the invocation pattern and observe relevant cold starts. The system is set up using the Kubeless framework and the RL environment is modelled for the agent to examine the necessary metrics to make guided decisions in provisioning an appropriate number of function instances.

We present an evidence of leveraging Q-Learning to address cold starts on FaaS platforms and verify it with improved platform throughput, reduced resource wastage while maintaining expected thresholds, during the iterations. We evaluate the performance of our proposed agent against the HPA policy and function queue policy. We successfully observe the RL-based agent outperforming comparing techniques after a training of 500 epochs that verifies our hypothesis of strong association between success rate and reduced number of cold starts on the platform. After the test analyses, the Q-Learning agent successfully improves throughput by up to 8.81\% and reduces resource wastage by up to 37\% while preparing sufficient functions to reduce cold starts.

\bibliographystyle{splncs04}
\bibliography{citation}
\end{document}